# Superconducting and normal phases of FeSe single crystals at high pressure

D Braithwaite[1], B Salce[1], G Lapertot[1], F Bourdarot[1], C Marin[1], D Aoki[1] and M Hanfland[2].

[1]*INAC / SPSMS, CEA-Grenoble, 17 rue des martyrs, 38054 Grenoble, FRANCE.*
[2]*ESRF, 6 Rue Jules Horowitz 38043 BP 220 Grenoble Cedex, France.*

**Abstract.** We report on the synthesis of superconducting single crystals of FeSe, and their characterization by X-ray diffraction, magnetization and resistivity. We have performed ac susceptibility measurements under high pressure in a hydrostatic liquid argon medium up to 14 GPa and we find that $T_C$ increases up to 33-36 K in all samples, but with slightly different pressure dependences on different samples. Above 12 GPa no traces of superconductivity are found in any sample. We have also performed a room temperature high pressure X-ray diffraction study up to 12 GPa on a powder sample, and we find that between 8.5 GPa and 12 GPa, the tetragonal PbO structure undergoes a structural transition to a hexagonal structure. This transition results in a volume decrease of about 16%, and is accompanied by the appearance of an intermediate, probably orthorhombic phase.

I Introduction

In the flurry of activity on the iron-arsenide superconductors, the discovery of superconductivity in tetragonal FeSe[1] is possibly an important piece in the puzzle of the mechanism of superconductivity in these systems. This is particularly true since a recent study[2] showed an enormous pressure effect where $T_C$ was found to increase from a relatively modest 8K at ambient pressure to 27K with a pressure of 2GPa. Very recently further studies find that $T_C$ reaches 36-37K at higher pressure[3-6].The fact that $T_C$ can reach such high values implies that the mechanism of superconductivity is probably similar to that of the Fe-As family. Furthermore, if this strong pressure effect can be understood, this might provide strong clues to understanding superconductivity in this system, and in the other iron based superconductors This increase of $T_C$ has been linked to with the enhancement of spin fluctuations[7] and with the reduction of the interlayer spacing[3].

So far all these studies are made on polycrystalline samples and in most cases in quasi-hydrostatic conditions, and there are serious discrepancies between the different results. The understanding of the system clearly needs the establishment of the intrinsic diagram, and this requires pure materials, and good hydrostatic conditions. The aim of this study is to determine the superconducting phase diagram of single crystals under pressure with argon as the hydrostatic pressure medium using an ac susceptibility measurement, and to compare the phase diagram with the structural changes found by X-ray diffraction studies.

II Sample preparation and characterization

Single crystals with typical dimensions 500 x 500 x 30 μm$^3$ were prepared by slow cooling of pre-reacted Fe (5N) and Se (6N) in an NaCl/KCl flux, with a nominal composition FeSe$_{0.85}$ similar to the procedure used by Zhang et al.[8]. Following a recent detailed analysis[9] of the metallurgical phase diagram of FeSe it is likely that the composition of the crystals, or at least of their superconducting volume, is in fact close to stoichiometric FeSe 1:1. The structure was checked by performing X-ray diffraction on about 20 crystals ground to powder. The X-ray spectrum (figure 1) shows a mix of the tetragonal PbO structure and hexagonal NiAs structure phases. The lattice parameters of the tetragonal phase (a = 3.7728 Å, c = 5.5233 Å) are in good agreement with the literature.

Crystals were checked for superconductivity by resistivity and magnetization measurements performed in commercial devices (Quantum design PPMS and MPMS) as shown in figure 2. Most crystals show a broad superconducting transition in the resistivity, with an onset at about 12 K and reaching zero resistivity at about 8 K. This is however considerably sharper than the transitions previously reported for flux grown crystals[8], and similar to results found in vapour transport grown crystals[10]. Measurements in the MPMS on single crystals were performed with a relatively high field (50 Oe) in order to obtain sufficient sensitivity. These show a significant decrease in the susceptibility both in field cooled and zero field cooled measurements although the overall susceptibility remains positive, possibly because of some ferromagnetic impurity. This does however imply that a substantial volume of the crystal is superconducting. The resistive transition was measured in magnetic field for both directions (inset figure 2 top) and shows a moderate anisotropy, with initial slopes of 2.8 T/K (H//c) and 4.0 T/K (H//ab).

III High pressure experimental procedures

Pieces of the crystals showing a superconducting transition were cut for measurements under pressure. The ac susceptibility was measured in a diamond anvil cell using a technique similar to that described in [11] where a pick-up coil consisting of about 10 turns of 12μm diameter copper wire is inserted in the sample chamber. The primary coil placed outside the sample chamber produced an excitation field of about 1 Oe at a frequency of 733 Hz. The pressure transmitting medium was argon. This medium provides good hydrostatic conditions at least up to about 10 GPa. The pressure was measured by the ruby fluorescence technique in-situ at low temperature with an accuracy of better than 0.1 GPa. A piece of lead was also placed inside the pick-up coil. This served the dual purpose of providing a second pressure measurement, and checking that the pick-up coil was functioning correctly. In order to compare the superconducting phase diagram with changes to the crystal structure, X-ray diffraction measurements under pressure were performed on a powder sample. This was prepared with a nominal composition $FeSe_{0.977}$, in order to be within the supposed optimum superconducting domain[9]. This sample also showed a mix of the tetragonal and hexagonal phases with a proportion of about 77:23. The powder sample was placed inside the hole (initial diameter 250 μm) of a stainless steel gasket of a membrane driven diamond anvil cell (DAC, culet size 600 μm) together with He as pressure transmitting medium and a ~ 5 μm diameter ruby sphere for pressure calibration. Diffraction images were collected at the ID09A beamline of the ESRF with a monochromatic beam (λ ~ 0.41 Å) focused to ~ 20 x 20 μm using a MAR555 flat panel detector. During exposure the cells were rotated by ±3° to improve powder averaging. Images were integrated with Fit2d software[12], and the structural analysis was performed using FullProf software[13].

IV Results and discussion

The ac susceptibility under pressure was performed on a total of 3 crystals. In figure 3 we show typical transitions found on one sample. The highest pressure where superconductivity was found was 10.5 GPa. In all cases no traces of superconductivity were found at pressures of 12 GPa and above, as can be seen in the inset of figure 3, where the detection of the Pb superconducting transition indicates that the pick-up coil is still intact. In figure 4 the phase diagram is shown for the different samples taking the onset criterion. We find that the pressure dependence of $T_C$ is somewhat sample dependent, but qualitatively similar for all samples. $T_C$ systematically reaches a maximum of 33–36 K at an optimum pressure of 6-8 GPa. At higher pressure $T_C$ decreases, and rapidly disappears. Sample 2 was measured twice, in the first run it was taken to 10.4 GPa where the onset temperature decreased to about 23K, then is was depressurized at room temperature and reloaded in a new pressure cell. On the second run the increase of $T_C$ was far less pronounced. The implications of this will be discussed later. The transitions were generally rather broad. It is difficult to define an objective criterion for the full width, so we have rather compared the half width defined as $\Delta T = T_{Conset} - T_{50\%}$. The relative transition width is physically more meaningful than the absolute value, so in order to characterize the transition width in figure 4b we have plotted the value $\Delta T / T_C$. The pressure dependence of this value is consistent for all samples, showing initially an increase with a maximum relative width at about 3 GPa, followed by a decrease to 8 GPa where quite sharp transitions are found. The relative width then increases very sharply again at higher pressures, except for the second run of sample 2.

The X-ray spectra at various pressures are shown in figure 5. We confirm the initial high compressibility ($B_0$ = 32 GPa) and the reduction of the ratio c/a with pressure found elsewhere[5, 14]. Up to about 8.5 GPa, lines from both the tetragonal PbO and hexagonal NiAs phases are present allowing us to follow the pressure dependence of the volume of both phases. Between 8.5 GPa and 10.5 GPa the tetragonal phase lines disappear and a more complicated structure is found, which can best be indexed as a mixture of the hexagonal and tetragonal phases, with an additional orthorhombic (Cmcm) phase. This orthorhombic phase can be understood as a necessary intermediate phase to allow the tetragonal to hexagonal transition accompanied by a large (16%) volume decrease. Above 10.5 GPa a hexagonal phase, with broad diffraction peaks is found. This may corresponds to a hexagonal NiAs phase structure with a distribution of cell parameters.. Our phase diagram is in agreement with the studies of Margadonna et al[3]. and Medvedev et al[5]. We do not find the high pressure orthorhombic phase seen by Garbarino et al[6]. though this is not necessarily in contradiction as it is reported to appear above 12 GPa, higher than the maximum pressure reached here.

We turn now to the discussion of the superconducting phase diagram. The maximum $T_C$ of 36-37 K has now been confirmed by 6 independent measurements and seems to be a robust feature of the system. However apart from this there are some discrepancies found in the phase diagram from one measurement to another. The non-anisotropic compressibility probably plays an important role on the change of $T_C$ as has already been pointed out. This would mean that the hydrostaticity of the pressure conditions might be crucial. All the previous studies have used solid pressure transmitting media, so only quasi-hydrostatic conditions, and this could partly explain some of the discrepancies. However this cannot be the only reason as in the results presented here we find a considerable difference in the pressure dependence of 2 crystals from the same batch, measured in the same conditions, with good hydrostaticity provided by the argon pressure transmitting medium. This implies that small differences in composition might produce quite different pressure dependences, while all samples still tend towards the same optimum value of $T_C$. At low pressures, p < 3GPa we found a broadening of the transition, reaching a maximum for the relative width at about 3 GPa. A similar feature was reported [15] and ascribed to a second superconducting phase due to the presence of a spin density wave. We certainly expect that sample inhomogeneity is the cause for this broadening, but we cannot at present distinguish between the existence of two phases, and just a spread of $T_C$ due to a continuous inhomogeneity of the sample. More interesting is the fact that in all samples we find much sharper transitions at higher pressures, with a minimum of the relative width at about 8 GPa in all samples, whether this corresponds to the optimum $T_C$ or not. This implies that at this pressure the sample is in a much more homogeneous state. The sudden increase in the transition width as pressure is increased above 8-9 GPa strongly suggests that sample inhomogeneity is increasing again, and this pressure corresponds to the onset of the structural transition found in this work and other studies[3, 5].

So far most of the high pressure structural studies have been performed only at room temperature, so it is delicate to compare the superconducting phase diagram with the observed structural changes, however the structural study by Margadonna et al[3], performed at 16 K, found the transition to the hexagonal phase at 9 GPa, precisely where this broadening occurs. Interestingly, for the sample that was pressurized a second time, the maximum $T_C$ was only 25 K, and no broadening of the transition above 8 GPa was seen. This could suggest that the high pressure phase achieved in the first run was retained even though pressure was completely released and the sample was heated to room temperature. Although the tetragonal to hexagonal transition has been shown to be reversible on powder samples, it is perhaps not on crystals. This would also imply that this phase is also superconducting but with a lower optimum critical temperature.

Considering the different pressure conditions, and sample preparations, the differences between the different studies at pressures below 10 GPa are minor. However there are large discrepancies between the different studies are at higher pressure. In one study[5] $T_C$ was found to decrease moderately to 25K at 15GPa, and no superconductivity was found at 29 GPa. In another study[3] $T_C$ was found to decrease more strongly, but the sample still showed signs of superconductivity at 14 GPa, with $T_C$ = 6K. Finally in another case[6] $T_C$ was found to increase continuously up to 22 GPa. In contrast in our study, we see a very strong decrease of $T_C$, down to about 22 K at 10.5 GPa, and then no signs of superconductivity at higher pressures. Of course neither resistivity nor ac susceptibility are bulk probes of superconductivity. Resistive transitions in particular can be seen with a tiny fraction of superconducting volume. As in all the studies cited here the resistive transitions at high pressure are invariably extremely broad, and in fact zero resistance is rarely found, we suggest that these transitions are not intrinsic, but due to a small fraction of the low pressure phase remaining in the sample, and this is probably emphasized by the use of non-hydrostatic conditions. In our case the use of the hydrostatic argon transmitting medium, and ac susceptibility which is a slightly better bulk probe than resistivity could explain why we see no traces of superconductivity. It seems to be a general feature that when hydrostatic conditions and bulk probes are used, the domains of existence of pressure induced superconductivity tends to shrink, as seen for example in heavy fermion superconductors[16], and this is probably what is happening here.

Pressure effects on the Fe-pnictide superconductors have been reviewed recently[17]. In this broader context, although the maximum $T_C$ reached here is well below the maximum of 55 K found in the optimum doped 1111 system[18], the value of 36 K found here is similar to the maximum $T_C$ of 38 K found in the 122 systems under pressure[19, 20], which is also the highest value found for any undoped Fe-pnictide superconductor.

V Conclusion

The determination of the high pressure superconducting phase diagram of FeSe is clearly far from completely established. In this study the use of single crystals and hydrostatic conditions is an important step towards this, and may help to understand some of the discrepancies found in different experiments. We confirm the strong increase in $T_C$, with an optimum value of about 36 K. Above 10.5 GPa at room temperature we find a complete transformation into the hexagonal NiAs structure, we see no traces of superconductivity at higher pressures. However the crystals measured here are clearly still not homogeneous and strong sample dependences are found for crystals from the same batch. To completely solve this question a new generation of pure crystals is necessary.

Acknowledgments

This work is supported by the Agence Nationale de la Recherche through Contract No. ANR-06-BLAN-0220.

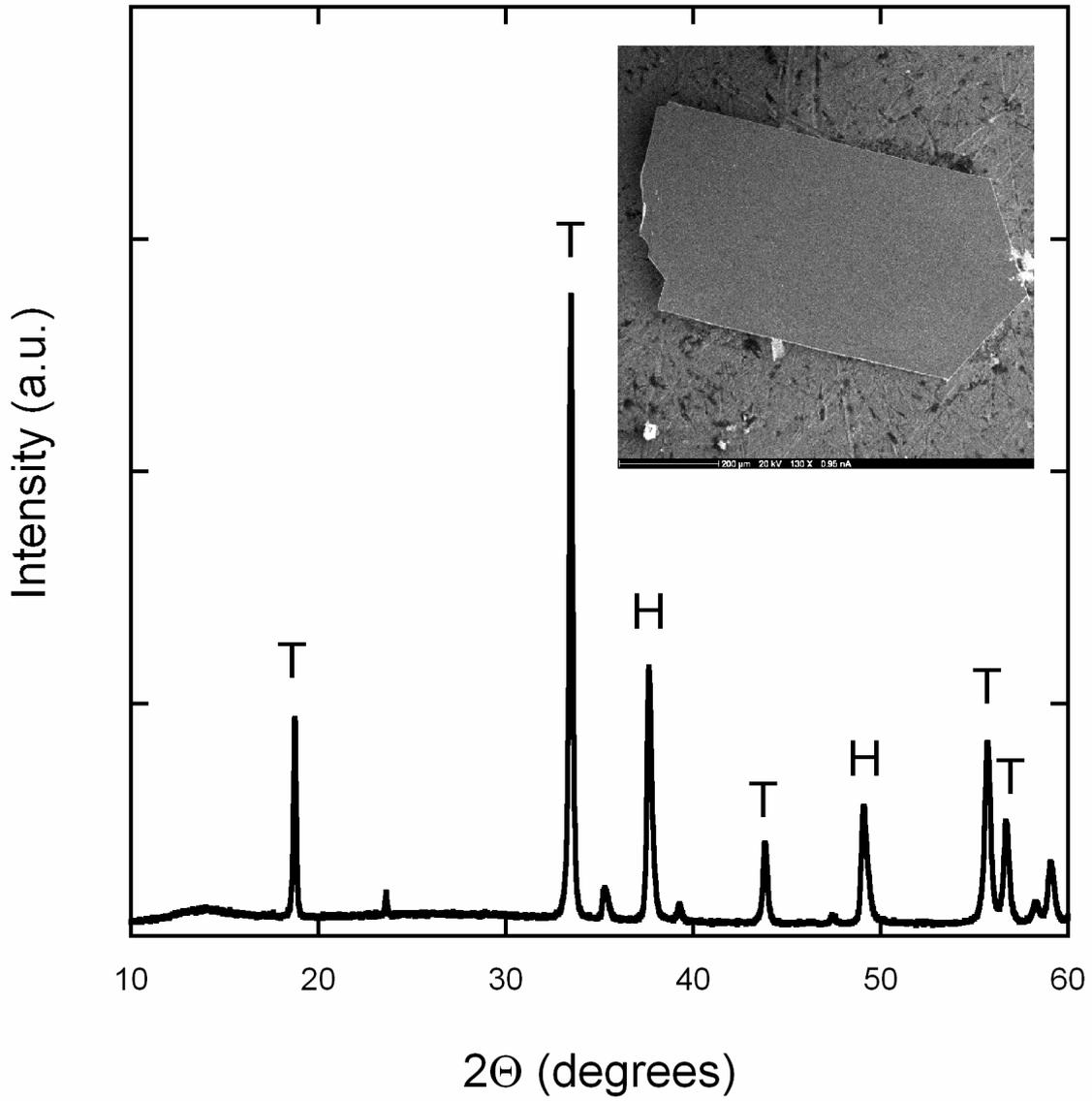

figure 1 : X-ray powder diffraction spectrum on about 20 single crystals ground to powder. T (respectively H) indicates the main peaks of the tetragonal PbO (respectively hexagonal NiAs) structure. Inset shows a SEM image of a typical single crystal.

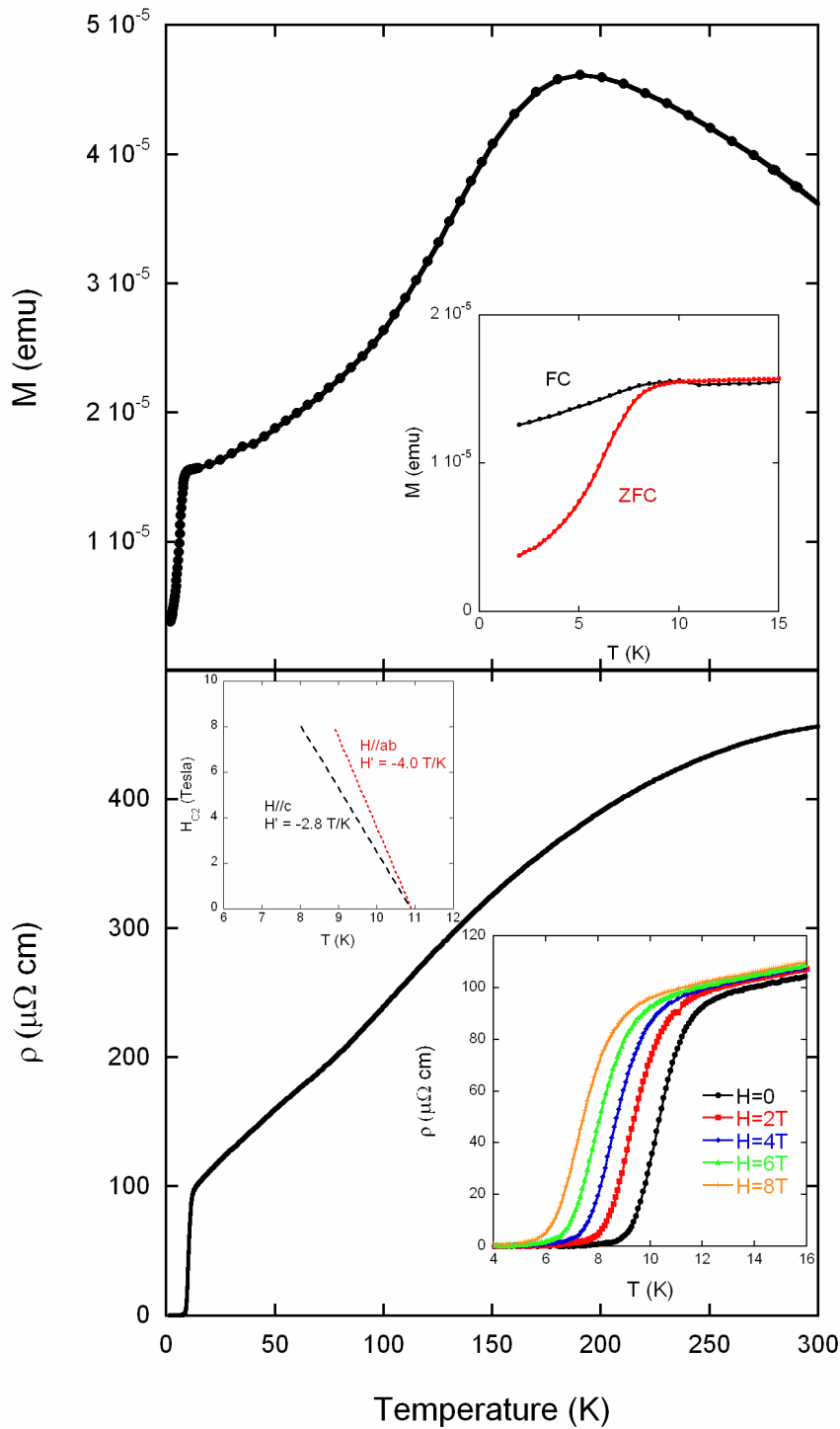

figure 2 : Magnetization (top) of a single crystal measured with a 50 oe field. Inset shows comparison of field cooled and zero field cooled measurements. Resistivity (bottom) on a single crystal of the same batch. Lower right inset shows transitions measured in a magnetic field H//c. Top left insert shows $H_{C2}$ measured for both orientations showing the anisotropy of the initial slope.

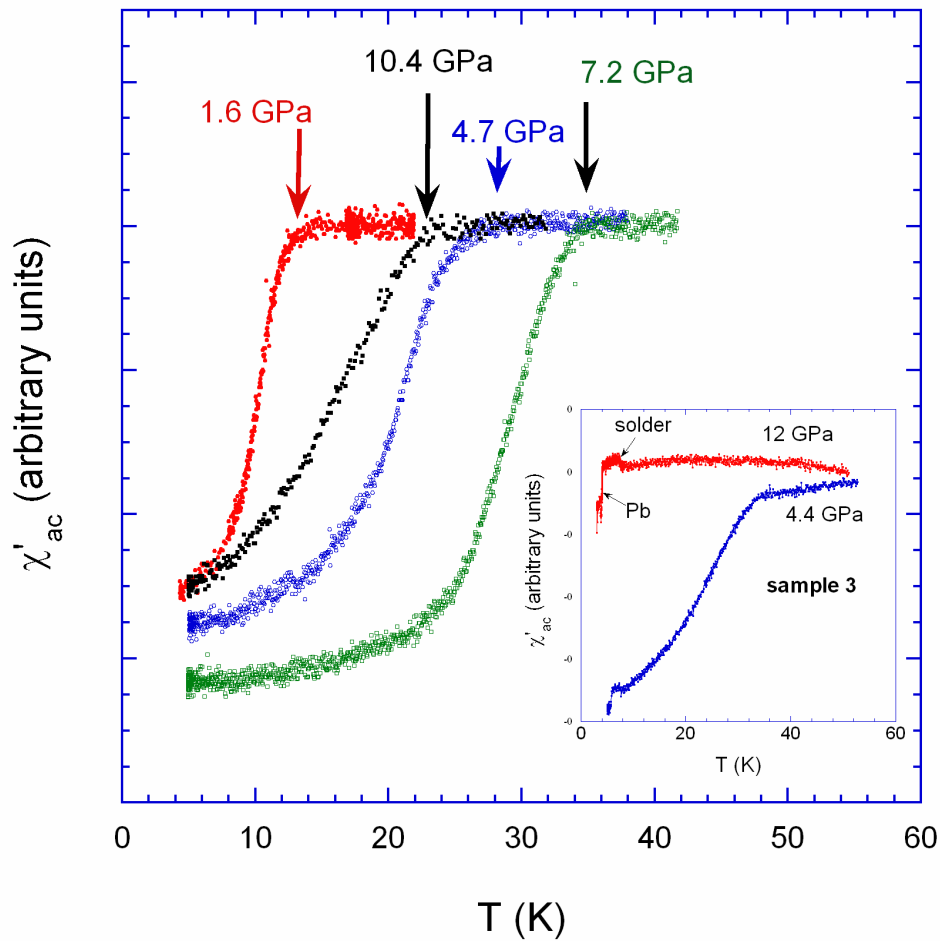

figure 3 : Superconducting transitions measured by ac susceptibility in the diamond anvil cell. No sign of superconductivity was found above 12 GPa in any of the 3 samples measured. The inset shows the comparison between 4.4 GPa and 12 GPa on sample 3. The continuous observation of the transition due to a small piece of lead placed inside the sample chamber proves that the detection coil was functioning correctly. The small anomaly at about 7K is due to soldered connections outside the cell.

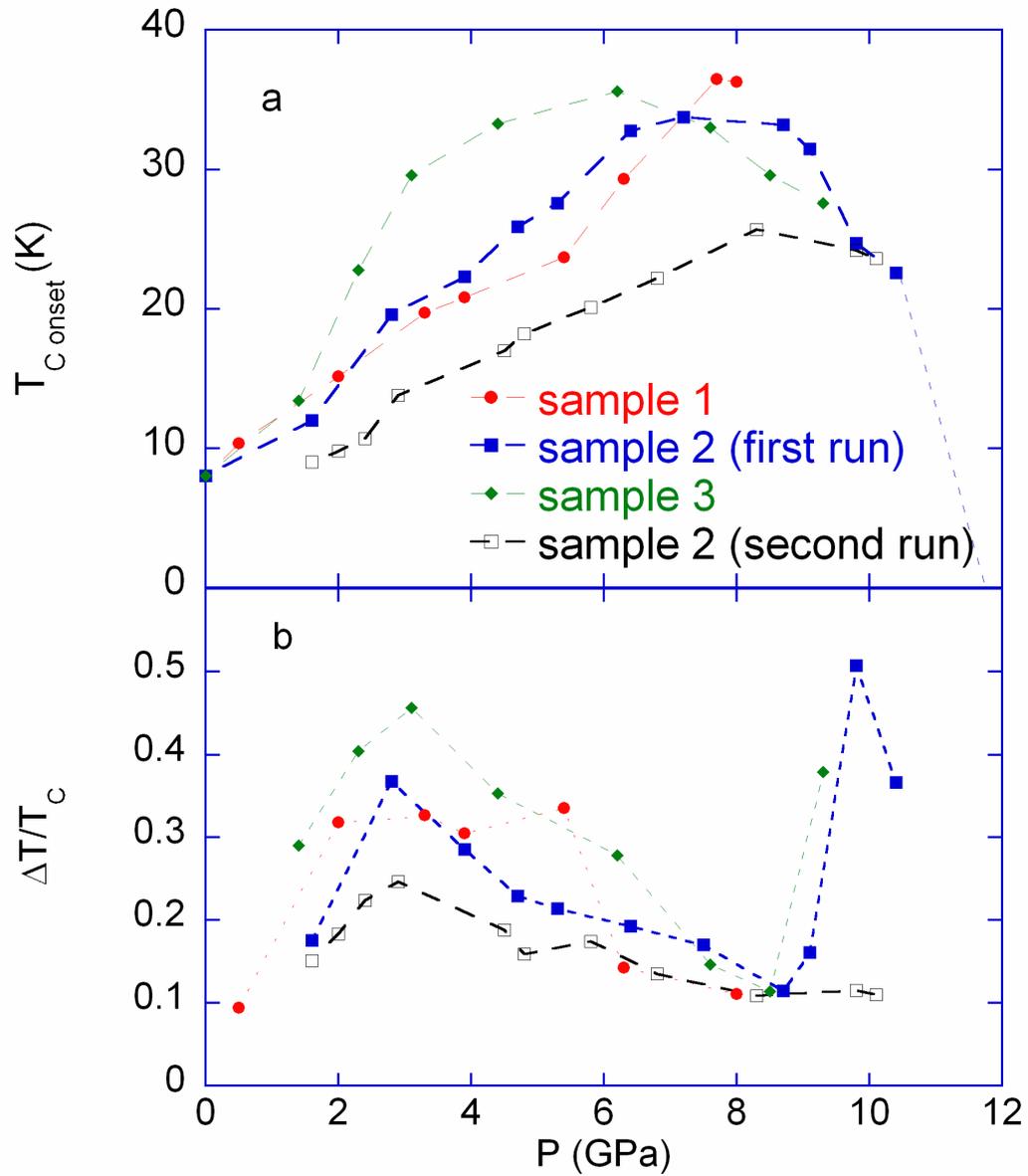

figure 4 : Superconducting phase diagram (top) obtained for different samples. The open squares correspond to the second run on sample 2 after total depressurization and heating to room temperature. Figure 4b (bottom) shows the evolution of the relative half width of the transition determined as $(T_{C\,onset} - T_{C50\%})/T_C$.

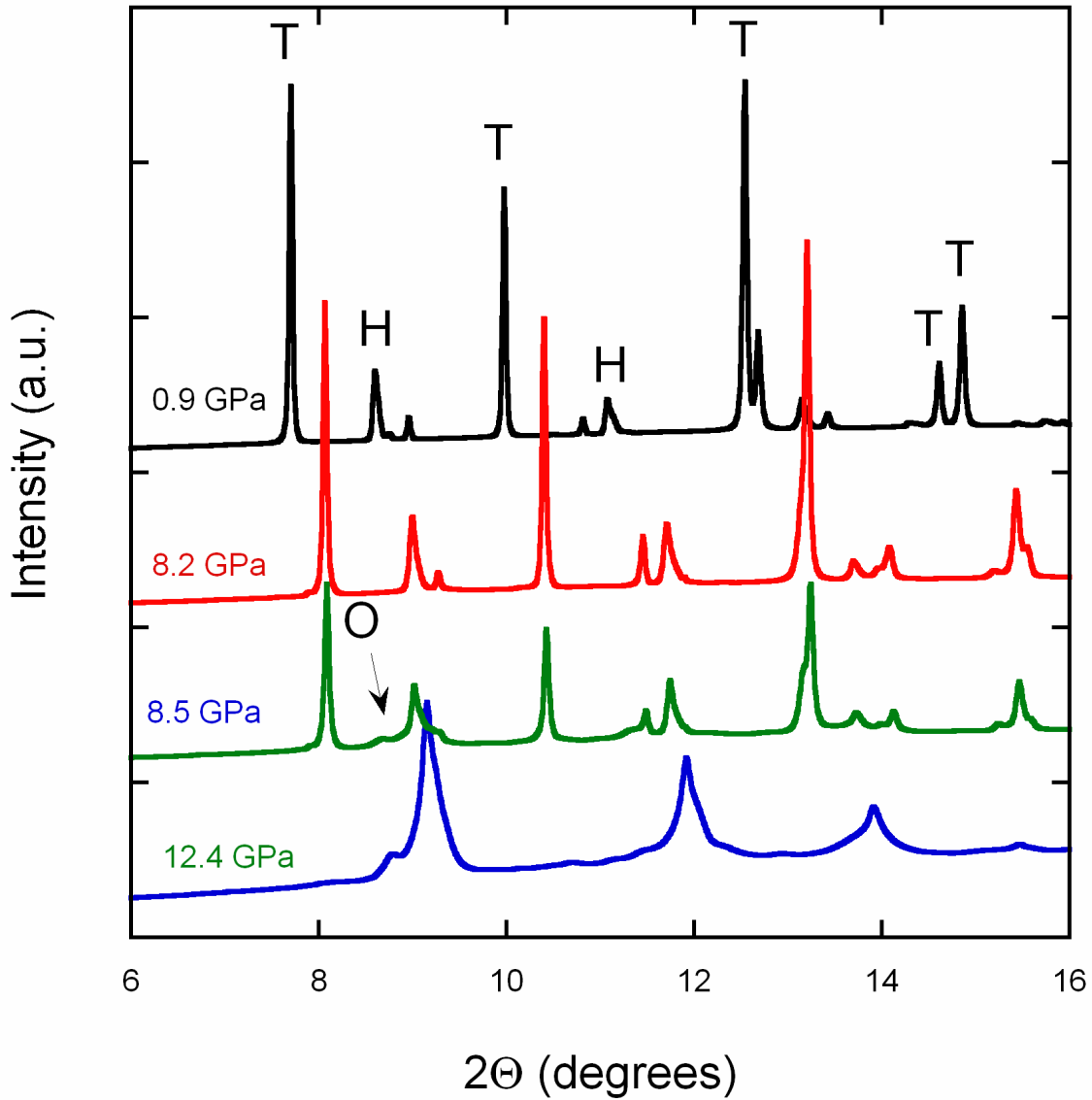

Figure 5 : Evolution of the X-ray spectra with pressure on a powder FeSe$_{0.977}$ sample. The low pressure spectrum shows a mixture of the tetragonal (T) phase and the hexagonal (H) phase. At 8.5 GPa the weight of the tetragonal phase is significantly decreased, with some extra features appearing (O) which can be indexed as an orthorhombic phase (Cmcm). At higher pressures the tetragonal phase disappears, and we index the spectrum as a hexagonal system with broad peaks which couls correspond to a distribution of lattice parameters.

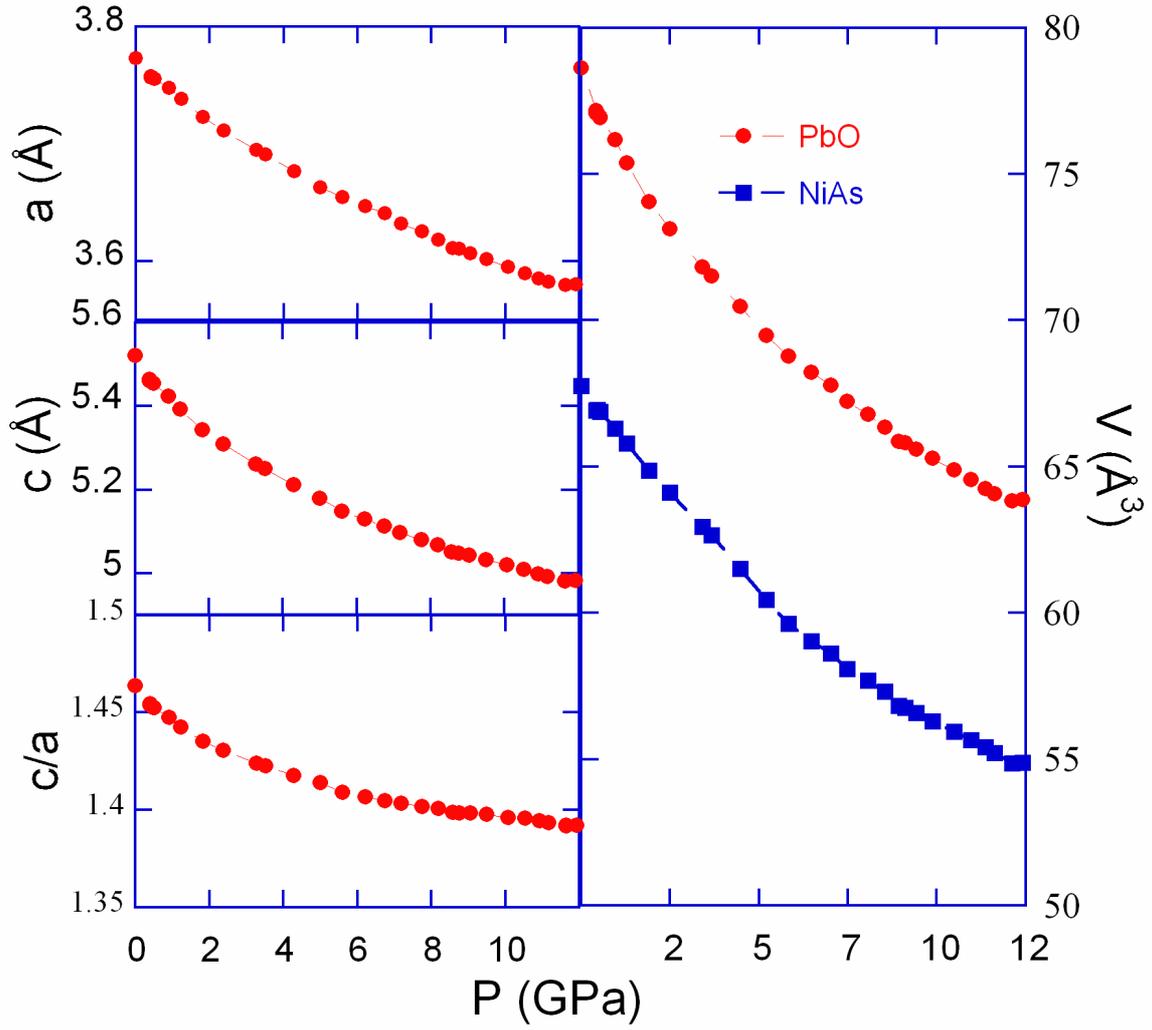

figure 6 : Evolution with pressure of the lattice parameters and the c/a ratio of the tetragonal (PbO) phase (left) and of the volume of both the tetragonal and hexagonal (NiAs) phases (right) on a FeSe$_{0.977}$ powder sample.